\documentstyle[preprint,aps,epsfig]{revtex}
\begin{document}
\title{$\pi \pi$ Scattering and the Meson Resonance Spectrum}
\author{W.M. Kloet}
\address{Department of Physics and Astronomy, Rutgers University,
PO Box 849, Piscataway, New Jersey 08855-0849, USA}
\author{B. Loiseau}
\address{Division de Physique Th\'eorique \footnote[1]{Unit\'e de 
Recherche des Universit\'es Paris 11 et Paris 6 Associ\'ee au CNRS},
Institut de Physique Nucl\'eaire, F-91406, Orsay CEDEX\\
and LPTPE, Universit\'e P. \& M. Curie, 4 Place Jussieu, 75252, Paris
CEDEX 05, France}
\date{\today}
\maketitle
\begin{abstract}
A $\pi \pi$, $\bar{K}K$, and $\rho \rho$($\omega \omega$) 
fully coupled channel model is used to predict the lowest isospin 
S, P, D, F-wave phase 
shifts and inelasticities for elastic $\pi \pi$ scattering 
from threshold to 2.0 GeV. 
As input the S-matrix is required to exhibit poles corresponding to 
the meson resonance table of the Particle Data Group. 
As expected, the $\pi \pi$ inelasticity is very strongly related to 
the opening of the $\bar{K}K$ channel near 1 GeV, and the opening of
$\rho \rho$(4$\pi$) and $\omega \omega$(6$\pi$) channels 
in the 1.5 GeV region. The predictions of this model are compared 
to the various 
elastic $\pi \pi \rightarrow \pi \pi$ amplitudes, 
that were obtained from analyses of 
$\pi^-$ p $\rightarrow$ $\pi^- \pi^+$ n data. 
The role of the various resonances, in particular the glueball 
candidate $f_0(1500)$ and the $f_J(1710)$ is investigated. 

\end{abstract}

\narrowtext
\tightenlines

\section{INTRODUCTION}

In heavy ion  collisions and in anti-nucleon physics  many pions are produced
in the final state.  Pion-pion scattering therefore plays an important role
in the final state interaction of these processes. 
Our knowledge of $\pi \pi$ scattering is
incomplete, in particular above $M_{\pi \pi} \approx$ 1~GeV. The dynamics of
$\pi \pi$ scattering is often described by effective meson-exchange in the
t-channel. 
The mechanism of t-channel exchange works very well in, 
for example, nucleon-nucleon scattering. 
However, for $\pi \pi$ scattering one may also consider the presence of 
the many resonances in the
s-channel, a feature that is typical for $\pi \pi$ scattering but 
not for nucleon-nucleon scattering. 
A list of relevant meson resonances with their properties can be 
obtained from the compilation of the Particle Data Group~\cite{PDG}. 
For energies larger than 1 GeV, an additional 
aspect is the strong coupling of the $\pi \pi$ channel 
to $\bar{K}K$ and multi-pion channels, 
for example, four and six pions.

Previous models for S-wave $\pi \pi$ scattering  have been applied
from threshold to 1.4 GeV.  
For example, separable  potential models have been
considered in an approach~\cite{can1,can2,can3,Kam1} 
in which the major focus was to obtain 
the proper low energy  behavior of the $\pi \pi$ S-wave phase shift, 
using the right combination of attraction and repulsion in the 
diagonal $\pi \pi$ interaction, 
and also to better understand the structure of the 
meson resonance $f_0$(980). 
More sophisticated one-meson exchange models~\cite{Loh,Mul,Pea} 
have been used up to 2 GeV, still within the framework of 
coupling $\pi \pi$ and $\bar{K} K$ channels.

In an earlier paper~\cite{Klo}  
we reported on results within a coupled three-channel model 
for S-wave scattering in the 1.0 -- 2.0 GeV energy region, 
and we were encouraged by its results when compared to 
the then available S-wave data. 
The relevant experimental $\pi \pi$ phase shifts and inelasticities, 
specifically $\pi^+ \pi^- $, can be found from the
analysis  by Protopopescu et al~\cite{Pro}, Grayer et al~\cite{Gra}, 
Hyams et al~\cite{Hya}, 
Bugg et al~\cite{Bug} and Kaminsky et al~\cite{Kam2}. 
These groups have extracted the pion-pion scattering amplitudes in 
the $M_{\pi \pi} = $ 0.60 -- 1.78 GeV region by obtaining them from 
an analysis of the reaction $\pi^-$ p $\rightarrow \pi^- \pi^+$ n. 
Kaminsky et al included also data from the same reaction with 
a polarized proton target, which allowed them to separate the
contributions due to $\pi$ and $a_1$ exchange. 
The low energy $\pi \pi$ data were published by Ref.~\cite{Ros}. 
From all these extracted data for the reaction 
$\pi^+ \pi^- $ $\rightarrow$ $\pi^+ \pi^- $ 
it is apparent that there is a rather
strong  energy dependence of the phase shifts 
as well as the inelasticities 
for S, P, D and F-wave $\pi^+ \pi^-$ elastic scattering. 
We do not know of any data above 1.78 GeV, 
so a model reproducing the existing 
data may eventually also serve as a basis for extrapolation to 
higher energies. 
Such an extrapolation would be required to study for example 
the final state interaction in the process 
$\bar{p}p \rightarrow \pi^+ \pi^-$~\cite{Bat}. 

In view of the presence of an extensive set of resonances 
as given by the Particle Data Group~\cite{PDG}, 
we have  opted in this paper for inclusion of the various 
known resonances in our model for $\pi \pi$ scattering. 
One can then address the question, 
what is the role of these resonances for elastic $\pi \pi$ scattering 
in the various partial waves in the region of 0.3 -- 2.0 GeV. 
In the energy range up to 2.0 GeV the dominant channels are 
$\pi \pi$, $\bar{K}K$, $\rho \rho$ and $\omega
\omega$. Within our model, as no spin effects have been taken into account,
the $\rho \rho$ and $\omega \omega$ channels can be described by a single
effective channel, e.g. $\rho \rho$. 

For example, the set of S-wave  resonances that play a role for 
the $\pi \pi$ channel are the $f_0$(980) and $f_0$(1370) resonances, 
but one should also include the recently discovered $f_0$(1500). 
While many conjectures about 
the nature of the $f_0$(1500) have been made, its role for $\pi \pi$ 
scattering is still an open question, and therefore 
will be one of the questions addressed in this paper. 
Another question is the $J$ value of the reported $f_J(1710)$ resonance. 
So far one only agrees that $J$ should be even~\cite{PDG}. 
In this approach one can study the implications 
of the presence of an isoscalar $f_0(1710)$ resonance
for S-wave $\pi \pi$ scattering, as well as 
the implications of the existence of an 
isoscalar $f_2(1710)$ resonance for D-wave $\pi \pi$ scattering. 
In practice we will impose the condition that the scattering matrix, 
corresponding to a particular angular momentum, should have poles at 
the complex energies of known resonances with appropriate quantum 
numbers and, from this input, one can study the implications for 
$\pi \pi$ scattering. 

The model that we will describe, can be applied for any angular 
momentum. In this paper it will be applied for S, P, D and F-waves. 
For each angular momentum the important input will be the 
location of the corresponding resonances as given by the 
Particle Data Group~\cite{PDG}. 
Sections II and III describe the model and show the analytic
expressions of the S-matrix for the various angular momenta. 
In sections IV and V  the results for respectively 
$\it{l}$ = 0 and $\it{l}$ = 2 are discussed. This order of angular 
momenta is chosen because the resonance $f_J(1710)$ may either 
contribute to the S-wave or D-wave scattering.  
Then the results for $\it{l}$ = 1 and $\it{l}$ = 3 are described 
in sections VI and VII respectively. 
Finally a discussion of all results and conclusions follows in section VIII.

\section{MODEL}

As in Ref.~\cite{Klo} we consider three coupled channels of 
$\pi \pi$, $\bar{K}K$ and $\rho \rho$, and label them respectively 
with the channel index $\it i$ = 1, 2, 3. 
Our method of derivation of the scattering amplitude 
is generalized to all angular momenta. 
For each angular momentum $\it{l}$ the Lippmann-Schwinger equation for the 
$\it{ij}$-element of the T-matrix at total invariant energy squared s, is 

\begin{equation}
<p|T_{ij}^l(s)|q> = <p|V_{ij}^l|q> - 
\frac{1}{2 \pi^2} \sum^3_{k=1} \int dp' p'^2 <p|V_{ik}^l|p'> G_k(p',s) 
<p'|T_{kj}^l(s)|q>.\label{lipp}
\end{equation}
Here $G_k(p',s)$ is the propagator for channel $\it{k}$ and $V_{ij}^l$ 
is part of the potential. 
In this case 
the potential as well as the T-matrix are both 3x3 matrices. 

We choose a separable form of the potential matrix elements 
for each angular momentum~$\it{l}$

\begin{equation}
<p|V_{ij}^l|q> = g_i^l(p) \lambda_{ij}^l g_j^l(q).\label{Vdef}  
\end{equation}
As a consequence the T-matrix elements for each $\it{l}$ 
are also separable 

\begin{equation}
<p|T_{ij}^l(s)|q> = g_i^l(p) \tau_{ij}^l(s) g_j^l(q),\label{Tdef}  
\end{equation}
where $\tau_{ij}^l$ satisfies

\begin{equation}
\tau_{ij}^l(s) = \lambda_{ij}^l - \sum^3_{k=1} \lambda_{ik}^l X_k^l(s) 
\tau_{kj}^l(s).\label{tau}
\end{equation}
The functions $X_k^l$ are defined in terms of the propagator $G_k$ and 
vertex functions $g_k^l$ 

\begin{equation}
X_k^l (s) = \frac{1}{2 \pi^2} \int dp' p'^2 g_k^l (p') G_k (p',s) 
g_k^l(p'),\label{Xdef} 
\end{equation}
and the functions $\tau_{ij}^l(s)$ can be obtained in closed form. 

In the following we will suppress as much as possible the label 
$\it{l}$ of the angular momentum. 
For example if the matrix $A$ for a given angular momentum 
$\it{l}$ is defined by 

\begin{equation}
A_{ij} = \lambda_{ij} + \lambda_{ij} X_j,\label{Adef} 
\end{equation}
the element $\tau_{11}$ has the general form 

\begin{equation}
\tau_{11} = \frac{(1+\lambda_{22} X_2) (1+\lambda_{33} X_3) - 
\lambda_{23} X_3 \lambda_{32} X_2 }{det(A)}.\label{tau11}
\end{equation}
The elastic $\pi \pi$ scattering amplitude for angular momentum 
$\it{l}$ is then given by the corresponding element $T_{ij}$ with 
$\it{i}$ = 1 and $\it{j}$ = 1, of the T-matrix of Eq.~(\ref{Tdef}). 
The model has a resonance of angular momentum $\it{l}$ if the 
corresponding T-matrix has a pole in the complex energy plane 
at an energy whose real part is the resonance mass and whose 
imaginary part is half the resonance total width. 
Such a pole corresponds to a zero of $det(A)$ at that complex 
energy, as can be seen from Eq.~(\ref{tau11}). 
The full expression for $det(A)$ is 

\begin{eqnarray}
det(A) = (1+\lambda_{11} X_1) (1+\lambda_{22} X_2) (1+\lambda_{33} X_3)  
-\lambda_{23} X_3 \lambda_{32} X_2 (1+\lambda_{11} X_1) \nonumber\\ 
-\lambda_{12} X_2 \lambda_{21} X_1 (1+\lambda_{33} X_3) 
-\lambda_{13} X_3 \lambda_{31} X_1 (1+\lambda_{22} X_2) \nonumber\\  
+(\lambda_{13} \lambda_{32} \lambda_{21}  
+\lambda_{12} \lambda_{23} \lambda_{31}) X_1 X_2 X_3,\label{detdef} 
\end{eqnarray}
with the couplings $\lambda_{ij}^l$ satisfying 

\begin{equation}
\lambda_{ji}^l = \lambda_{ij}^l. 
\end{equation}

We choose the form factor $g_i^l(p)$ of Eq.~(\ref{Vdef}) to be 

\begin{equation}
g_i^l(\beta,p) = \sqrt{\frac{4 \pi}{m_i}} \frac{p^l}{(\beta^2+p^2)^{l+1}}, 
\label{gdef}
\end{equation}
where $\beta$ carries an index $\it{i}$ as well as $\it{l}$. 
For the propagator $G_i(p',s)$ we take the form 

\begin{equation}
G_i(p',s) = \frac{m_i}{p'^2-p_i^2-i\epsilon} ,
\label{Gdef}
\end{equation}
where s is related to $p_i$ by 

\begin{equation}
s=4(p_i^2+m_i^2).
\end{equation}
The functions $X_i^l$ for $\it{l}$ = 0 to 3, which are dependent 
on the range parameters $\beta_i^l$, become in this case 

\begin{equation}
X_i^{l=0}(\beta,p,s) = \frac{1}{2 \beta (\beta - i p)^2 } ,
\label{X0def}
\end{equation}

\begin{equation}
X_i^{l=1}(\beta,p,s) = \frac{\beta^2 - 4 i \beta p - p^2}
{16 \beta^3 (\beta - i p)^4 } ,
\label{X1def}
\end{equation}

\begin{eqnarray}
X_i^{l=2}(\beta,p,s) = \frac{3 \beta^4 - 18 i \beta^3 p  - 
38 \beta^2 p^2 + 18 i \beta p^3 + 3 p^4} 
{256 \beta^5 (\beta - i p)^6 } ,
\label{X2def}
\end{eqnarray}

\begin{eqnarray}
X_i^{l=3}(\beta,p,s) = \frac{5 \beta^6 - 40 i \beta^5 p  - 
131 \beta^4 p^2 + 208 i \beta^3 p^3 
+131 \beta^2 p^4 -40 i \beta p^5 - 5 p^6} 
{2048 \beta^7 (\beta - i p)^8 } .
\label{X3def}
\end{eqnarray}
Again the indices $\it{i}$ and $\it{l}$ of $\beta$ and 
$\it{i}$ of $\it{p}$ are suppressed. The above expressions follow from 
evaluation of the integral in Eq.~(\ref{Xdef}) for real values of 
$\it{p}$ using the definitions Eq.~(\ref{gdef}) and Eq.~(\ref{Gdef}). 
Details of the evaluation are given in part A of the Appendix. 
The integral in Eq.~(\ref{Xdef}) is defined for Re($\it{p}$) $\geq$ 0, and 
is dis-continuous over the unitarity cut for positive real $\it{p}$. 
Therefore it cannot be analytically continued for complex $\it{p}$ 
with Im($\it{p}$) $<$ 0. However, the expressions of 
Eqs.~(\ref{X0def}),~(\ref{X1def}),~(\ref{X2def}),~(\ref{X3def}) 
are continuous and analytic for all complex momenta $\it{p}$, 
except for a pole on the negative imaginary axis, 
and can therefore be used as analytic continuation 
of $X_i^l(\beta,p,s)$ to complex values of $\it{p}$, even to the 
resonance region where Im($\it{p}$) $<$ 0. 
One also notes the familiar property of $X_i^l(\beta,p,s)$ 

\begin{equation}
X_i^l(\beta,p,s) = [X_i^l(\beta,-p^*,s^*)]^*.
\label{Xstar}
\end{equation}

As mentioned previously, 
the channels $\it i$ = 1, 2, 3 correspond to $\pi \pi$, $\bar{K} K$ and 
$\rho \rho$ respectively, and we use $m_1$ = 0.1396 GeV, 
$m_2$ = 0.4937 GeV and $m_3$ = 0.7680 GeV. 
Since no spin aspects are considered and the mass 
of the $\omega$ meson is very close to the mass 
of the $\rho$ meson, the $\rho \rho$ channel can effectively be viewed as 
representing also the $\omega \omega$ channel for angular momenta 
where I = 0. 
We ignore the width of the $\rho$ meson, and assume that 4$\pi$ and 
6$\pi$ channels are dominated 
by $\rho \rho$ and $\omega \omega$ respectively. 
The relative momenta, $p_i$, in the $i^{th}$ channel, are related by the
kinematic condition

\begin{equation}
\frac{s}{4} = \frac{1}{4} M^2_{\pi \pi} = 
p^2_1+m^2_1 = p^2_2+m^2_2 = p^2_3+m^2_3.
\end{equation}
This equation defines the invariant mass $M_{\pi\pi}$. 
For $\it{l}$ = 0 the form factor in coordinate space has the familiar form
\begin{equation}
g_i(r) = \frac{e^{- \beta_i r}}{r} .  \label{ff}
\end{equation}

The above method is equivalent to previously given
expressions~\cite{Klo} for the S-wave elastic $\pi \pi$ 
scattering amplitude, 

\begin{equation}
S_{11} = \frac{D(-p_1,p_2,p_3)}{D(p_1,p_2,p_3)},
\end{equation}
where $S_{11}$ is the $\pi \pi$ S-matrix element. $D$ is the Jost 
function, and $D$ = $det(A)$, i.e.

\begin{eqnarray}
D(p_1,p_2,p_3) = [( R_1 + \Lambda_1 ) (R_2 + \Lambda_2 ) (R_3 + \Lambda_3 ) 
- \Lambda^{2}_{12} (R_3 + \Lambda_3 ) 
- \Lambda^{2}_{13} (R_2 + \Lambda_2 ) \nonumber\\
- \Lambda^{2}_{23} (R_1 + \Lambda_1 ) 
+ 2 \Lambda_{12} \Lambda_{13} \Lambda_{23}]/(R_1 R_2 R_3).\label{Jost}
\end{eqnarray}
The couplings $\Lambda^l_{i}$ and $\Lambda^l_{ij}$ as well as the 
functions $R^l_i$ are dimensionless and 
their relation to $\lambda^l_{ij}$ and $X^l_i$ of Eq.~(\ref{detdef}) 
are given in 
Table~\ref{table:PARA}. 
Since the function $X_i^l$ has the property given in Eq.~(\ref{Xstar}), 
the Jost function satisfies 

\begin{equation}
D(p_1, p_2, p_3) = [D(-p_1^*, -p_2^*, -p_3^*)]^*.
\label{Dstar}
\end{equation}

For each angular momentum $\it{l}$ = $J$ we now require the S-matrix 
to have poles at the known $J^{\pi}$ resonances.
A summary of the considered resonances appears in 
Table~\ref{table:resonances}.
For example, for $J^{\pi}$ = 0$^+$ the resonances are 
$f_0$(980), $f_0$(1370) and $f_0$(1500). 
For each angular momentum $\it{l}$ = $J$ there are three range
parameters $\beta _i$, and six couplings $\lambda_{ij}$. 
The three poles are then  chosen at the  complex energies, 
(Re$M_{\pi \pi}$, Im$M_{\pi \pi}$) corresponding respectively to the mass 
of the resonance and half its total decay width. 
Again for $J^{\pi}$ = $0^+$, these energies would be 
respectively  (0.980, -0.0250), (1.300, -0.200) and (1.503, -0.060) GeV.
We require $det(A)$ of Eq.~(\ref{detdef}) to have zeros at these complex 
energies, 
which leads to six constraints on the nine parameters. There are several 
parameter sets that satisfy the above constraints, as will be discussed below.

\section{ADDITIONAL FORMALISM FOR $J^{\pi}$ = $0^+$, $I = 0$}

The case $J^{\pi}$ = $0^+$ needs special attention. 
We have before published
results~\cite{Klo} for a three coupled channel model 
as described in Section II. 
One obtains impressive results for phase shifts as well as 
inelasticities in the energy region 0.9 -- 2.0 GeV. 
However, within the model described in Section II 
it is not possible to obtain 
a realistic description at energies below 0.9 GeV. 
The reason is that the potential for the $\pi \pi$ channel for the 
S-channel requires an attractive term as well as a repulsive one. 
It has been earlier demonstrated~\cite{can2} that addition of a second 
term in the S-wave $\pi \pi $ potential is successful in describing 
the low energy behavior of the $\it{l}$ = 0 scattering. 
We develop therefore a formalism, described in Appendix B, adding 
such an attractive term. 
In this paper this formalism is only applied for $\it{l}$ = 0 scattering.

\section{RESULTS FOR $J^{\pi}$ = $0^+$, $I = 0$}

Since our previous publication of $\it{l}$ = 0 results~\cite{Klo}, 
the Particle Data Group~\cite{PDG} has significantly revised 
its compilation of $J^{\pi}(I)$ = $0^+(0)$ resonances. 
At present it contains the resonances 
$f_0$(980), $f_0$(1370), $f_0$(1500) and possibly $f_0$(1710). 
They occur respectively at 
the complex energies (0.980, -0.025), (1.300, -0.200), 
(1.503,-0.060) and (1.697, -0.088) GeV. 
However, the real as well as the imaginary part of the pole positions 
have uncertainties and those are rather large for the $f_0$(1370) resonance.  

We are particularly interested in the role played by the ``new'' 
resonance $f_0$(1500), which has been considered as a glueball candidate. 
Therefore we start by taking the first three resonances,  
$f_0$(980), $f_0$(1370) and $f_0$(1500), as input for our model and study 
the consequences for the resulting $\pi \pi$ scattering parameters. 
These three $J$ = 0 resonances together will impose constraints on the 
analytic expression for the S-matrix for $\it{l}$ = 0. The S-matrix 
must have three poles at the corresponding three complex energies. 
These constraints determine or restrict the possible 
values of the model parameters $\beta_0,\beta_1,\beta_2,\beta_3,
\Lambda_0, \Lambda_1,\Lambda_2,\Lambda_3,\Lambda_{12},\Lambda_{13}$ 
and $\Lambda_{23}$, where, as stated in Appendix B, the index 0 labels 
the second potential term of the $\pi\pi$ channel. 
Using these parameter values, one can then immediately determine the 
S-submatrix $S_{11}$, describing $\pi \pi \rightarrow \pi \pi $ 
scattering. 
Since $S_{11}$ for angular momentum $\it{l}$ 
is parametrized by a phase shift $\delta$ and an inelasticity $\eta$, 
viz. 

\begin{equation}
S_{11} = \eta e^{2 i \delta},\label{unitarity}
\end{equation}
one obtains the values of $\delta$ and $\eta$ as a function of energy. 
The $\it{l}$ = 0 scattering parameters 
are then compared to experimental data. 

There is still an additional freedom in our model. This 
is related to the location of the resonance in the complex energy plane 
and to the analytic structure of the S-matrix for 
the three coupled channels $\pi \pi$, $\bar{K}K$ and $\rho \rho$. 
Since the relation between energy and momentum is quadratic, for each 
complex resonance energy there are two corresponding complex momenta 
$p_i$ for each channel, one with Im~$p_i$ $<$ 0 and a second
with Im~$p_i$ $>$ 0. 
This defines two possible sheets in the complex energy plane for each 
channel. 
Because there are three channels, there are all together eight
sheets. 
The exact location off all resonance poles with respect to these 
eight sheets determines the specific solution for the S-matrix. 

In our standard choice of complex channel momenta $p_i$ 
for each resonance, 
the sign of Im~$p_i$ is such that the momenta $p_i$ are all 
close to the physical scattering region. 
The physical scattering region has 
positive real momenta $p_i$ for the open channels, 
and positive imaginary momenta $p_i$ for the closed channels. 
This means that a standard resonance corresponds to a channel 
momentum $p_i$ with Im~$p_i$ $<$ 0 if channel $\it i$ is open 
and Im~$p_i$ $>$ 0 if channel $\it i$ is closed. 
The $\pi \pi$ channel is always open, the $\bar{K}K$ channel opens 
at 2$m_{K}$ and the $\rho \rho$ channel opens at 2 $m_{\rho}$. 

\begin{figure}[htbp]
\centerline{\psfig{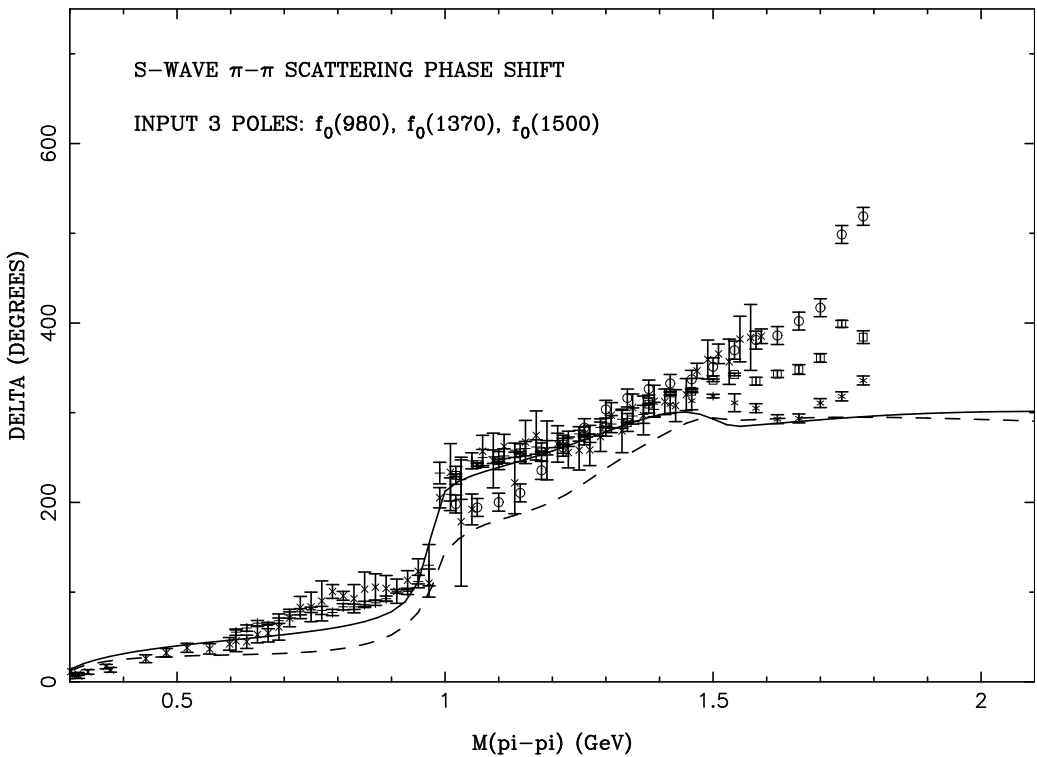}}  
\caption[]{
S-wave $\pi \pi$ phase shifts from the three-channel model, 
if poles are present for $f_0(980)$, $f_0(1370)$ and $f_0(1500)$. 
The dashed curve is the result for $\it{l}$ = 0 if the $f_0(1500)$ 
pole occurs at Im~$p_{\pi \pi}$ $<$ 0 
with the parameter set of the first line in 
Table~\ref{table:LVAL0}.
The solid curve is the prediction for the parameter set in the second 
line in Table~\ref{table:LVAL0}, 
where the $f_0(1500)$ resonance pole corresponds to 
Im~$p_{\pi \pi}$ $>$ 0. 
Data are from Refs.~\cite{Gra,Hya,Kam2}. }
\label{fig:figdel0PDG}
\end{figure}

If we put all three resonances 
$f_0$(980), $f_0$(1370) and $f_0$(1500), 
on the standard $\pi \pi$ , $\bar{K}K$ and $\rho \rho$ sheets, 
i.e. near the physical region, 
(where all three channel momenta $p_1, p_2, p_3$ have the required 
imaginary part, in agreement with each channel being open or closed),  
a solution is found for the parameter set ($\beta_i, \Lambda_i, 
\Lambda_{ij}$) that is given in the first line of 
Table.~\ref{table:LVAL0}. 
However, the corresponding prediction for the 
$\pi \pi$ phase shifts is considerably lower than the data 
suggests, as can be seen by the dashed curve in 
Fig.~\ref{fig:figdel0PDG}.  
Therefore we next consider cases where one of the resonances 
is not close to the physical scattering region. 

From the decay properties of the first three $J$ = 0 
resonances one knows that 
$f_0$(980) and $f_0$(1370) both have significant couplings to the 
$\pi \pi$ and $\bar{K}K$ channels. 
On the other hand the $f_0$(1500) resonance has a 
preference to decay into $\eta \eta'$, $\eta \eta$, 4$\pi^o$,
2$\pi^o$ and 2$\pi^+$2$\pi^-$. It is significant that 
the decay of $f_0$(1500) into $\pi^+ \pi^-$ and
$\bar{K}K$ may be rather small. 
It suggests that the $f_0$(1500) resonance could be located on the 
$\pi^+ \pi^-$ or $\bar{K}K$ non-standard sheets 
(i.e. its pole location may correspond to a 
channel momentum $p_i$ with a positive imaginary part). 

\begin{figure}[htbp]
\centerline{\psfig{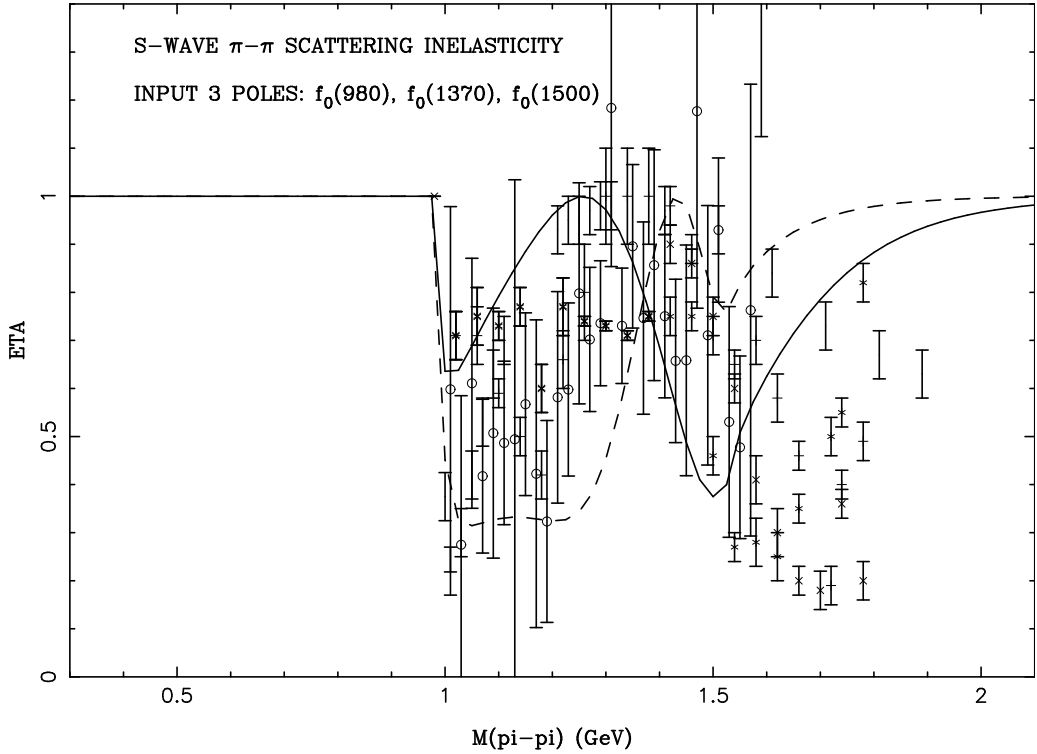}}  
\caption[]{
S-wave $\pi \pi$ inelasticities from the three-channel model, 
if poles are present for $f_0(980)$, $f_0(1370)$ and $f_0(1500)$. 
The dashed curve is the result for $\it{l}$ = 0 if $f_0(1500)$ 
has Im~$p_{\pi \pi}$ $<$ 0 and the parameters of the first line in 
Table~\ref{table:LVAL0}.
Solid curve is the prediction for the parameter set in the second 
line in Table~\ref{table:LVAL0}, 
where $f_0(1500)$ has Im~$p_{\pi \pi}$ $>$ 0. 
Data are from Refs.~\cite{Gra,Hya,Kam2}. }
\label{fig:figeta0PDG}
\end{figure}

If we choose to put $f_0$(1500) on the non-standard $\pi^+ \pi^-$ sheet, 
(i.e. treat it as a virtual resonance, where the relative $\pi \pi$ 
momentum has a positive imaginary part), there is a significant 
improvement of the model prediction when compared to the experimental data. 
The predictions for the $\pi \pi$ phase shifts and inelasticities
are shown as the solid curve in Fig.~\ref{fig:figdel0PDG} and 
Fig.~\ref{fig:figeta0PDG}. 
The corresponding parameter set is given in the second line of Table~
\ref{table:LVAL0}. 
The behavior of phase shift and inelasticity above 1.6 GeV will be 
influenced further by higher resonances. 
On the other hand, if we put the $f_0$(1500) resonance on the sheet, 
where also the relative $\bar{K}K$ momentum has a 
positive imaginary part, the $\pi \pi$ 
phase shift obtains a significant structure in the 1.2 -- 1.4 
GeV region. Such a structure is not present in the experimental data. 
We therefore conclude that the data for $\pi^+ \pi^-$ scattering 
are consistent with a $f_0(1500)$ resonance that is very different 
from the $f_0(980)$ and $f_0(1370)$ 
and that it couples only weakly to $\pi^+ \pi^-$. 
At the same time these data do not require its coupling to $\bar{K}K$ 
to be also weak.

\begin{figure}[htbp]
\centerline{\psfig{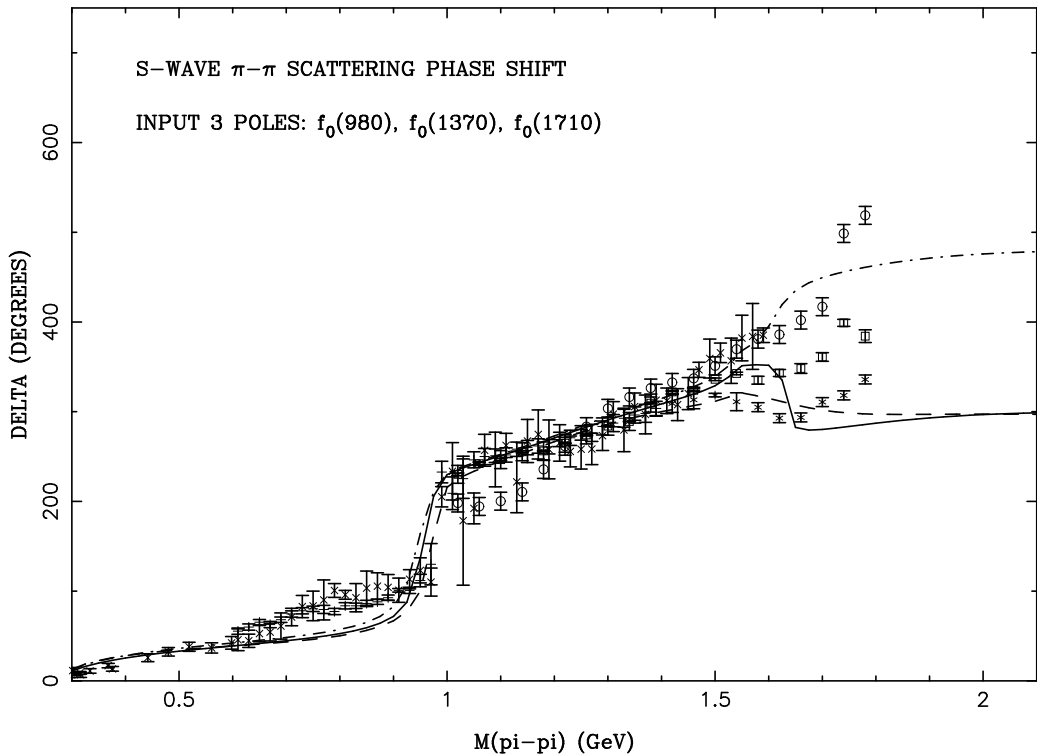}}  
\caption[]{
S-wave $\pi \pi$ phase shifts from the three-channel model, 
if poles are present for $f_0(980)$, $f_0(1370)$ and $f_0(1710)$. 
Solid, dashed and dash-dotted curves are the prediction for 
$\it{l}$ = 0 parameter sets from 
Table~\ref{table:LVAL0}, lines three thru five. 
Data are as in Fig.~\ref{fig:figdel0PDG}. }
\label{fig:figdel00}
\end{figure}

\begin{figure}[htbp]
\centerline{\psfig{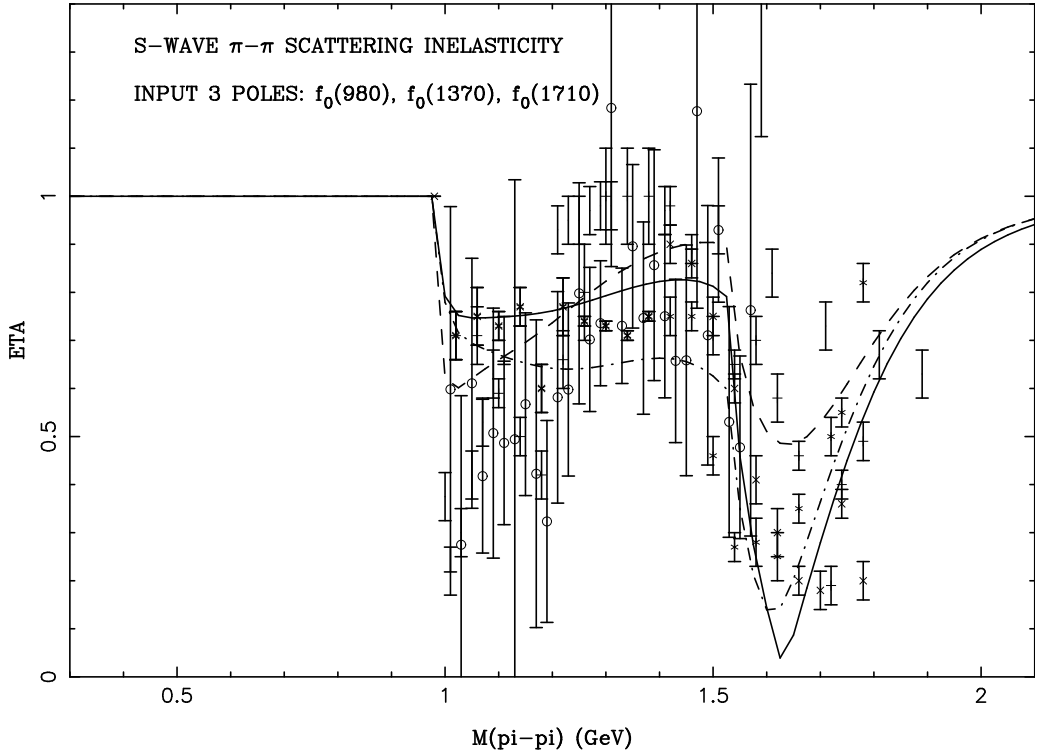}}  
\caption[]{
S-wave $\pi \pi$ inelasticities from the three-channel model, 
if poles are present for $f_0(980)$, $f_0(1370)$ and $f_0(1710)$. 
Curves are as in Fig.~\ref{fig:figdel00}. 
Data are as in Fig.~\ref{fig:figeta0PDG}. }
\label{fig:figeta00}
\end{figure}

The next question is if these data are consistent with the existence 
of a $f_0(1710)$ resonance. 
Indeed it is possible with resonance poles at the 
complex masses (1.200, -0.250) and (1.697, -0.300) to obtain an 
excellent prediction for the phase shift in the 
1.0 -- 1.5 GeV region that is in good agreement with the experimental 
data. In 
Fig.~\ref{fig:figdel00} and Fig.~\ref{fig:figeta00} the S-wave phase
shifts and S-wave inelasticities for three parameter sets, are shown 
when $f_0(1710)$ is included.
In this case one requires a large width for $f_0(1710)$ as 
well as a low mass for the $f_0(1370)$.  
The corresponding three sets of parameters are given in the third to 
fifth line of Table~\ref{table:LVAL0}. 
The phase shifts in this case (see Fig.~\ref{fig:figdel00}) show 
an energy dependence that is closer to the 
data than the model prediction in Fig.~\ref{fig:figdel0PDG}. 
This makes a strong case in favor of $J$ = 0 for $f_J(1710)$. 
The possibility of the existence of $f_2(1710)$ still remains to be 
considered and 
the results for $J$ = 2 will be discussed in the next section.

\section{RESULTS FOR $J^{\pi}$ = $2^+$, $I = 0$}

The Particle Data Group lists several $J^{\pi}(I)$ = $2^+(0)$ resonances, 
$f_2$(1270), $f_2$(1525) and $f_2$(2010) with the possibility of 
an additional $f_2$(1710). 
The first three resonances occur at complex energies 
(1.275, -0.092), (1.525, -0.038) and (2.011, -0.101) GeV. 
However, a resonance with even $J$ has also been 
observed at (1.697, -0.088) GeV and we will therefore follow two avenues. 
First, in case A we impose on our model the 
existence of three resonances in $J$ = 2 at complex energies 
(1.275, -0.092), (1.525, -0.038) and (1.607, -0.088) GeV, 
while in the second case B the existence of three resonances at 
(1.275, -0.092), (1.525, -0.038) and (2.011. -0.101) GeV 
will be used as a model constraint. We will subsequently discuss 
both cases. 

In both cases we use the formulae given in Section II. The 
S-submatrix $S_{11}$ for D-waves is then described by the nine 
parameters 
$\beta_1,\beta_2,\beta_3,\Lambda_1,\Lambda_2,\Lambda_3,\Lambda_{12},
\Lambda_{13}$ and $\Lambda_{23}$.
The requirement that the S-matrix must have poles at the complex 
energies corresponding to three resonances, 
leads to six constraints. However, 
it turns out that neither case A nor case B allows a solution within our 
model if we demand that all three resonances are on the standard sheet 
for $\pi \pi$, $\bar{K}K$ and $\rho \rho$ (i.e. where all relative channel 
momenta have imaginary parts close to the physical region). 
On the other hand solutions for both cases are found, if 
we locate the second resonance $f_2(1525)$ on the $\pi \pi$ 
sheet, where the relative $\pi \pi$ 
momentum has a positive imaginary part. 
A physical interpretation of this would be that the coupling of the 
$f_2(1525)$ resonance to the $\pi \pi$ channel is weak, meaning that 
$f_2(1525)$ has a rather small decay width into $\pi \pi$. 
The Particle Data Group gives that 
$f_2(1525)$ decays for only 0.8\% into $\pi \pi$ 
while for 88.8\% into $\bar{K}K$, which would justify to put the 
$f_2(1525)$ pole not on the standard $\pi \pi$ sheet near the physical 
region, but place it on the alternative $\pi \pi$ sheet, 
where Im $p_{\pi\pi}$ $>$ 0. 
Proceeding, the model gives results for case A as well as for case B 
closest to the data if the 
third resonance is put on the non-standard $\rho \rho$ sheet. 

\begin{figure}[htbp]
\centerline{\psfig{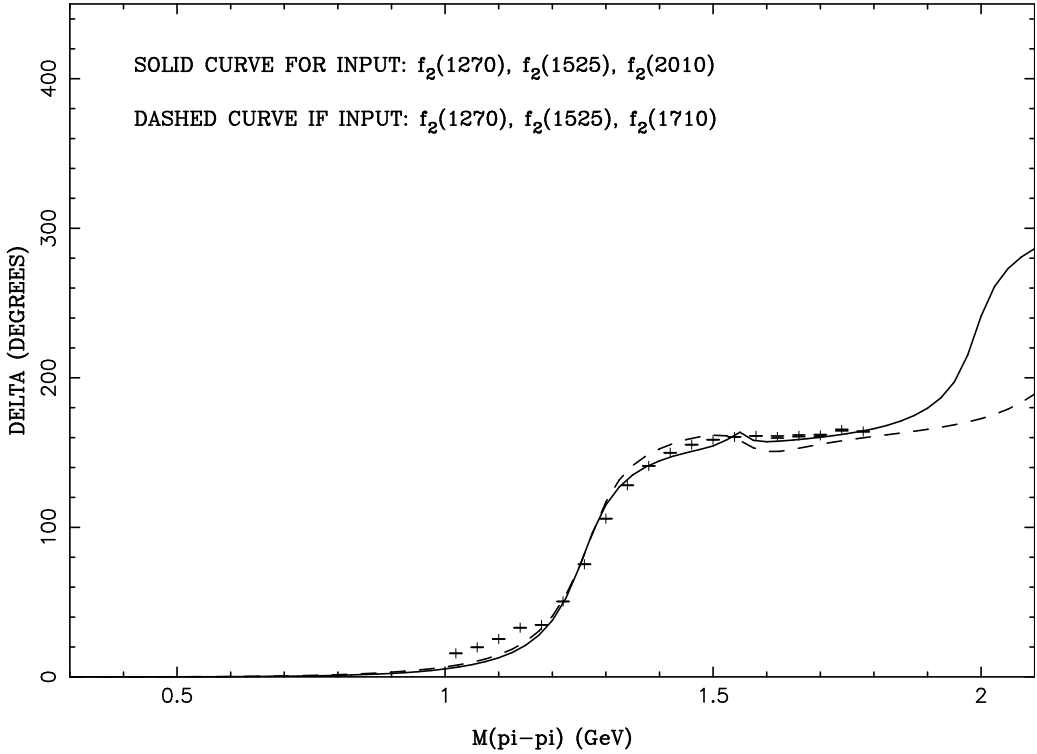}}  
\caption[]{
D-wave $\pi \pi$ phase shifts from the three-channel model, 
if poles are present for $f_2(1270)$, $f_2(1525)$ and $f_2(2010)$. 
Dashed curve is the prediction from the first $\it{l}$ = 2 parameter 
set from 
Table~\ref{table:LVAL2}, case A, 
with the $f_2(1525)$ resonance with Im~$p_{\pi \pi}$ $>$ 0, 
while the third resonance is the $f_2(1710)$. 
Solid curve is the prediction from the second $\it{l}$ = 2 parameter 
set from 
Table~\ref{table:LVAL2}, case B, 
with the $f_2(1525)$ resonance with Im~$p_{\pi \pi}$ $>$ 0,  
while the third resonance is the $f_2(2010)$. 
Data are from Ref.~\cite{Hya}. }
\label{fig:figdel2}
\end{figure}

\begin{figure}[htbp]
\centerline{\psfig{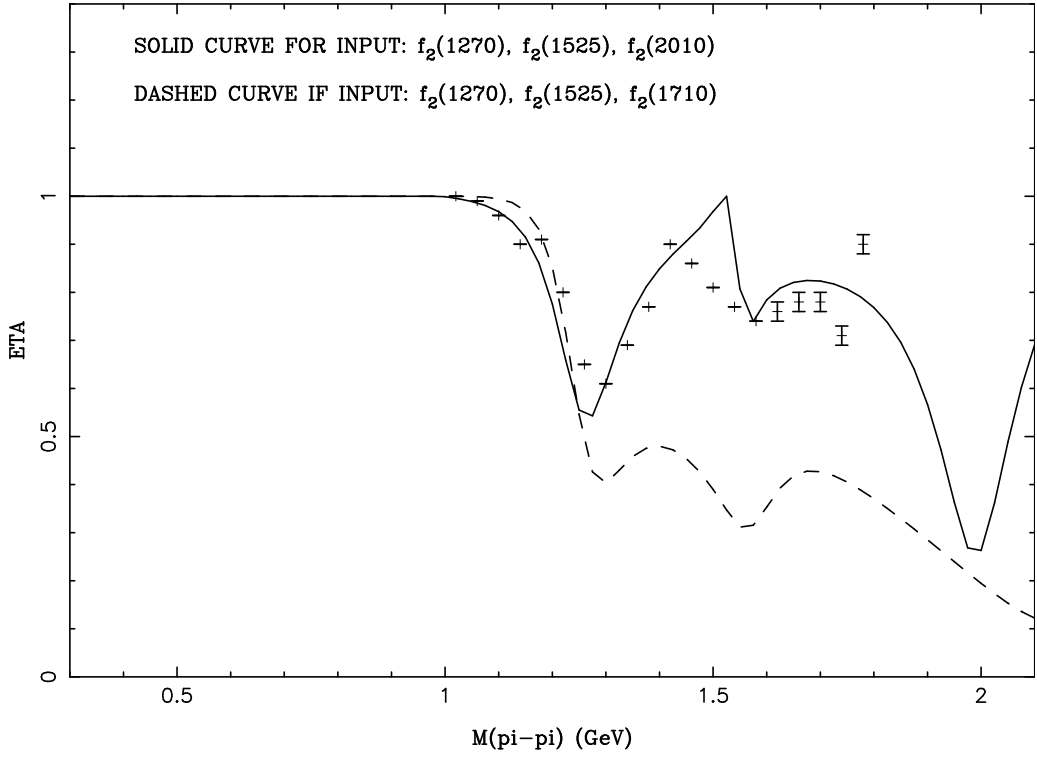}}  
\caption[]{
D-wave $\pi \pi$ inelasticities from the three-channel model, 
if poles are present for $f_2(1270)$, $f_2(1525)$ and $f_2(2010)$. 
Dashed and solid curves are as in Fig.~\ref{fig:figdel2}. 
Data are from Ref.~\cite{Hya}. }
\label{fig:figeta2}
\end{figure}

Then for case A, which contains the $f_2(1710)$ resonance, 
the parameter set is given in the first line of 
Table~\ref{table:LVAL2}, 
and the model predictions for the phase shift and inelasticity are 
given by the dashed line in Fig.~\ref{fig:figdel2} and 
Fig.~\ref{fig:figeta2} respectively.
For case A a structure near 1.6 GeV is present in the phase shift, 
and the inelasticity is much too low in comparison to the data. There 
is no evidence for a similar structure in the experimental phase shift. 

On the other hand, 
if we consider case B where there is no resonance at 1.7 GeV, but
instead a resonance near 2.0 GeV, the parameter set is given on 
the second line of Table~\ref{table:LVAL2}, 
and the corresponding model prediction 
for phase shift and inelasticity is represented by the solid curves 
in Fig.~\ref{fig:figdel2} and Fig.~\ref{fig:figeta2}. 
In the phase shift of case B the structure near 1.6 GeV has 
practically disappeared, 
and the prediction for the phase shift (see Fig. 5) 
is in closer agreement with the data. 
At the same time one finds that for case B the predicted 
inelasticity (see Fig. 6) is in much better agreement with the 
experimental inelasticity. 

It is tempting to conclude that the 
presence of a $f_2(1710)$ resonance would cause more structure in the 
$\it{l}$ = 2 phase shift than has been observed experimentally, 
and a value of $J$ = 2 for $f_J(1710)$ is therefore unlikely.

\section{RESULTS FOR $J^{\pi}$ = $1^-$, $I = 1$}

\begin{figure}[htbp]
\centerline{\epsfig{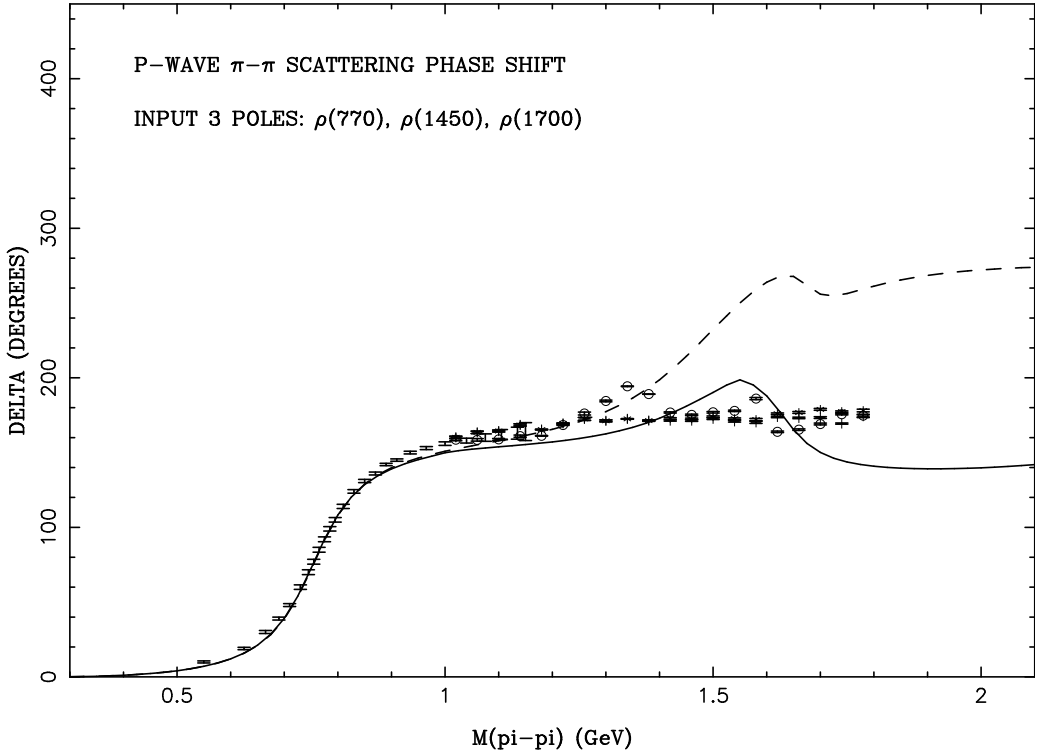}}  
\caption[]{
P-wave $\pi \pi$ phase shifts (in degrees) 
from various experimental analyses. 
The curves represent $\pi \pi$ phase shifts for $\it{l}$ = 1 from the 
three-channel model, 
if poles are present for 
$\rho(770)$, $\rho(1450)$ and $\rho(1700)$. 
The dashed curve has all resonances on the standard 
$\pi \pi$, $\bar{K}K$ and $\rho \rho$ sheets, while the solid curve 
is the result if $\rho (1700)$ is located on the sheet where 
Im~$p_{\bar{K}K}$ $>$ 0.  
Data points are from Refs.~\cite{Pro,Hya}. }
\label{fig:figdel1}
\end{figure}

\begin{figure}[htbp]
\centerline{\psfig{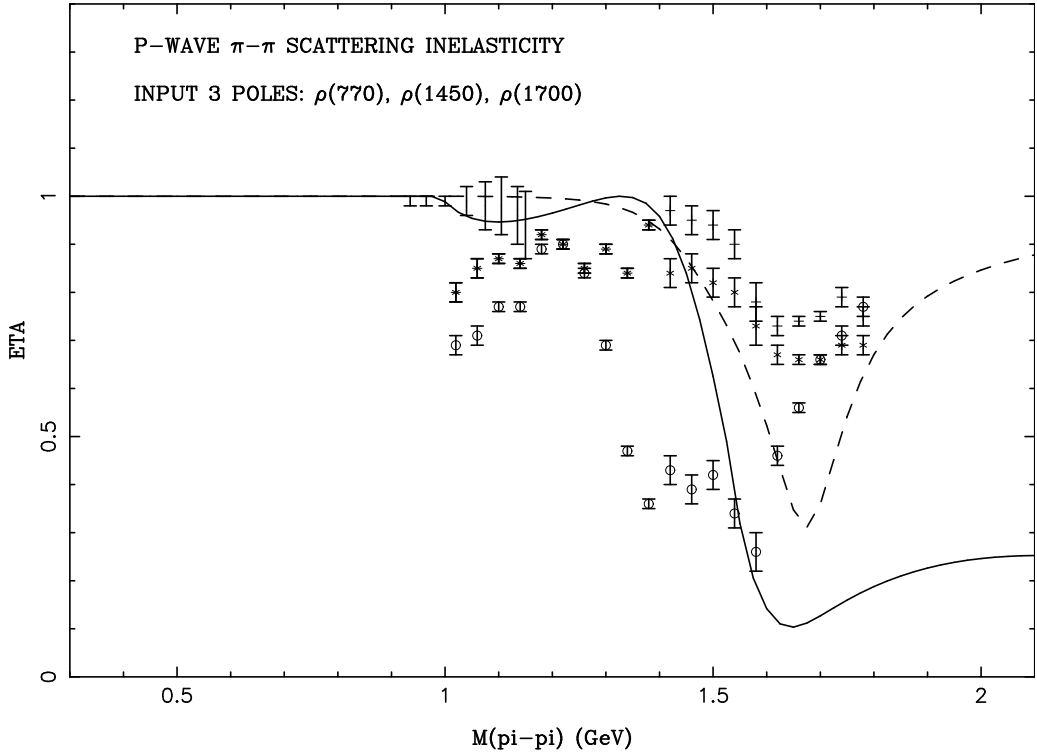}}  
\caption[]{
P-wave $\pi \pi$ inelasticities
from various experimental analyses. 
The curves represent $\pi \pi$ inelasticities for $\it{l}$ = 1 from the 
three-channel model, 
if poles are present for 
$\rho(770)$, $\rho(1450)$ and $\rho(1700)$. 
The dashed curve has all resonances on the standard 
$\pi \pi$, $\bar{K}K$ and $\rho \rho$ sheets, while the solid curve 
is the result if $\rho (1700)$ is located on the sheet where 
Im~$p_{\bar{K}K}$ $>$ 0.  
Data points are from Ref.~\cite{Pro,Hya}. }
\label{fig:figeta1}
\end{figure}

Following again the formalism of Section II, the resonances imposed for 
$\it{l}$ = 1
are $\rho$(770), $\rho$(1450) and $\rho$(1700). They occur,  
according to the most recent compilation of the Particle Data Group, 
at the complex energies (0.768, -0.075), (1.465, -0.,155) 
(1.700, -0.118) ~\cite{PDG} in units of GeV.  
Therefore the existence
of these three $J$ = 1 resonances again imposes the constraints that the 
analytic expression for the S-matrix for $\it{l}$ = 1 has three poles at these 
three complex energies. 
These constraints determine the possible 
values of the nine model parameters, 
and lead to corresponding P-wave phase shifts and inelasticities. 
The parameter values for our model are given 
in the first line of Table~\ref{table:LVAL1}. The resulting phase shifts and
inelasticities are shown respectively in Fig.~\ref{fig:figdel1} 
and Fig.~\ref{fig:figeta1} by the dashed curve. 
The different sets of experimental data for the $\it{l}$ = 1 phase shift 
are in quite good agreement with each other. 
One observes that the model phase shifts are in good agreement 
with the data only below 1.3 GeV, but are too high above 1.3 GeV. 
The model phase shift also shows a structure near 1.6 GeV that is
not present in the data. 
For the $\it{l}$ = 1 inelasticity there is no agreement between the 
different experimental data sets. 
In one experimental set of data there is considerable inelasticity 
when the $\bar{K}K$ channel opens at 1 GeV, while in others the
inelasticity is mainly driven by the opening of the $\rho \rho$ 
channel near 1.4 -- 1.5 GeV. 
Hence it is not possible to draw any conclusions about the model 
predictions from the inelasticity. 
The model result as seen in the dashed curve in 
Fig.~\ref{fig:figeta1} has only a contribution to the inelasticity 
beyond the 1.4 GeV region. 
It is somewhat of a surprise that a treatment of 
$\rho$(770), $\rho$(1450) and $\rho$(1700) 
as being very similar type resonances in all three channels, 
causes the model to fail for energies above 1.3 GeV. 

We have explored several of the options of placing one of the 
resonances on another sheet. 
The best agreement with the experimental data is obtained if 
we locate the third resonance, $\rho(1700)$,  on the non-standard 
$\bar{K}K$ sheet (i.e. Im~$p_{\bar{K}K}$ $>$ 0). 
For this case the corresponding parameter set is given in the second 
line of Table~\ref{table:LVAL1}, 
and the corresponding prediction of the phase shift and inelasticity is 
given as the solid curves in Fig.~\ref{fig:figdel1} and 
Fig.~\ref{fig:figeta1}. 
Comparing with the dashed curve, a dramatic improvement has been obtained 
for the phase shift. A structure near 1.6 -- 1.7 GeV remains however 
present in the model. 
There may be an indication of some structure in one set of 
experimental phase shifts, but it is far from compelling. 
The corresponding model prediction for the inelasticity includes 
for this case a larger effect of the $\bar{K}K$ channel, as shown by 
the solid curve in Fig.~\ref{fig:figeta1}. 
As long as the experimental 
data for the inelasticity suffer from the considerable internal 
disagreement as shown in Fig.~\ref{fig:figeta1}, it is hard to draw 
conclusive information from a comparison of this observable at this time. 
Other choices of possible sheets do not lead to further improvement 
of the predictions.

\section{RESULTS FOR $J^{\pi}$ = $3^-$, $I = 1$}

\begin{figure}[htbp]
\centerline{\psfig{figure=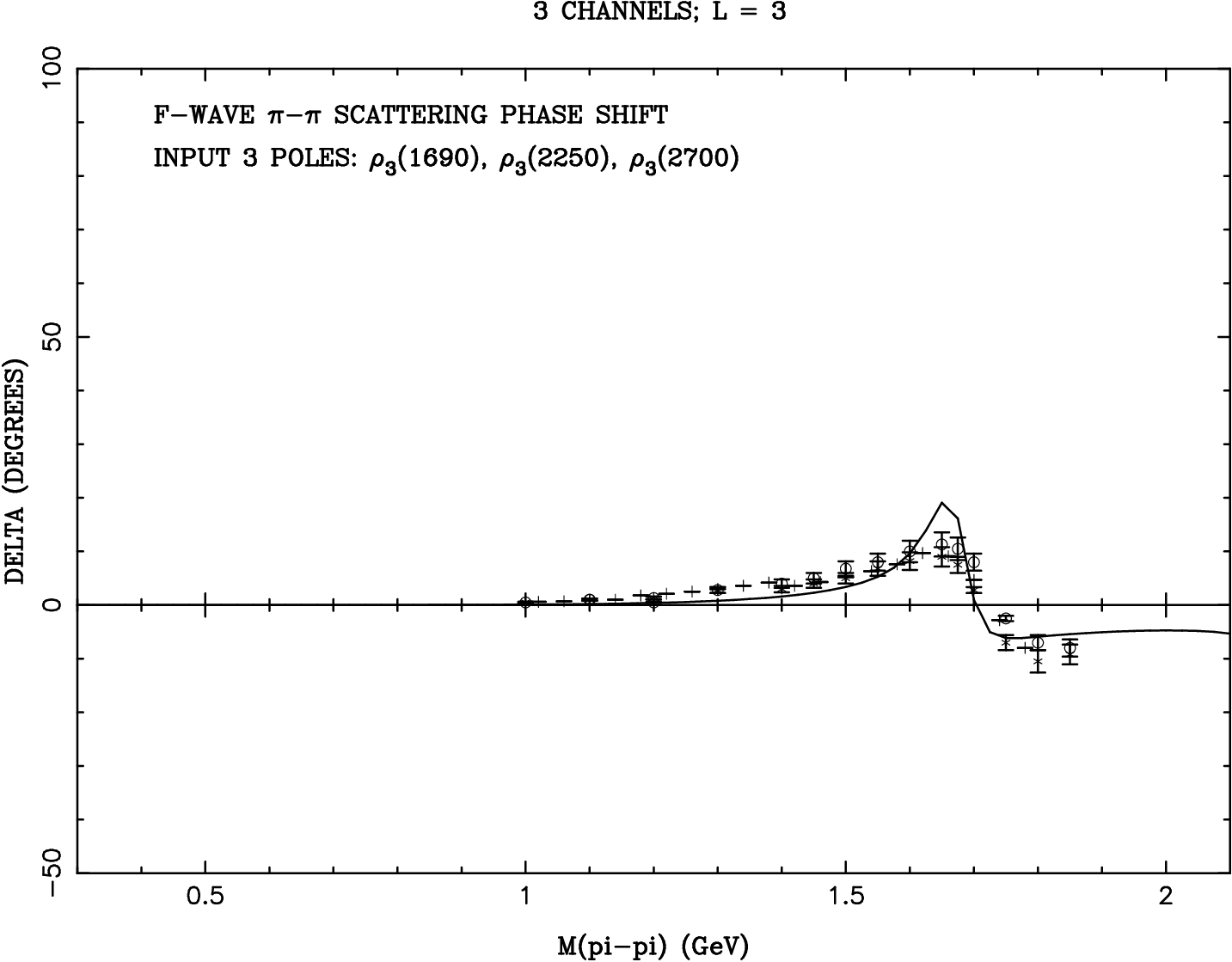,angle=-90,width=14cm
,bbllx=35pt,bblly=15pt,bburx=630pt,bbury=720pt,clip=}}  
\caption[]{
F-wave $\pi \pi$ phase shifts from the three-channel model, 
if poles are present for 
$\rho_3(1690)$, $\rho_3(2250)$ and $\rho_3(2700)$. 
Solid curve is the prediction for the $\it{l}$ = 3 parameter set from 
Table~\ref{table:LVAL3}. 
Data are from Refs.~\cite{Hya,Bug}.}
\label{fig:figdel3}
\end{figure}

\begin{figure}[htbp]
\centerline{\psfig{figure=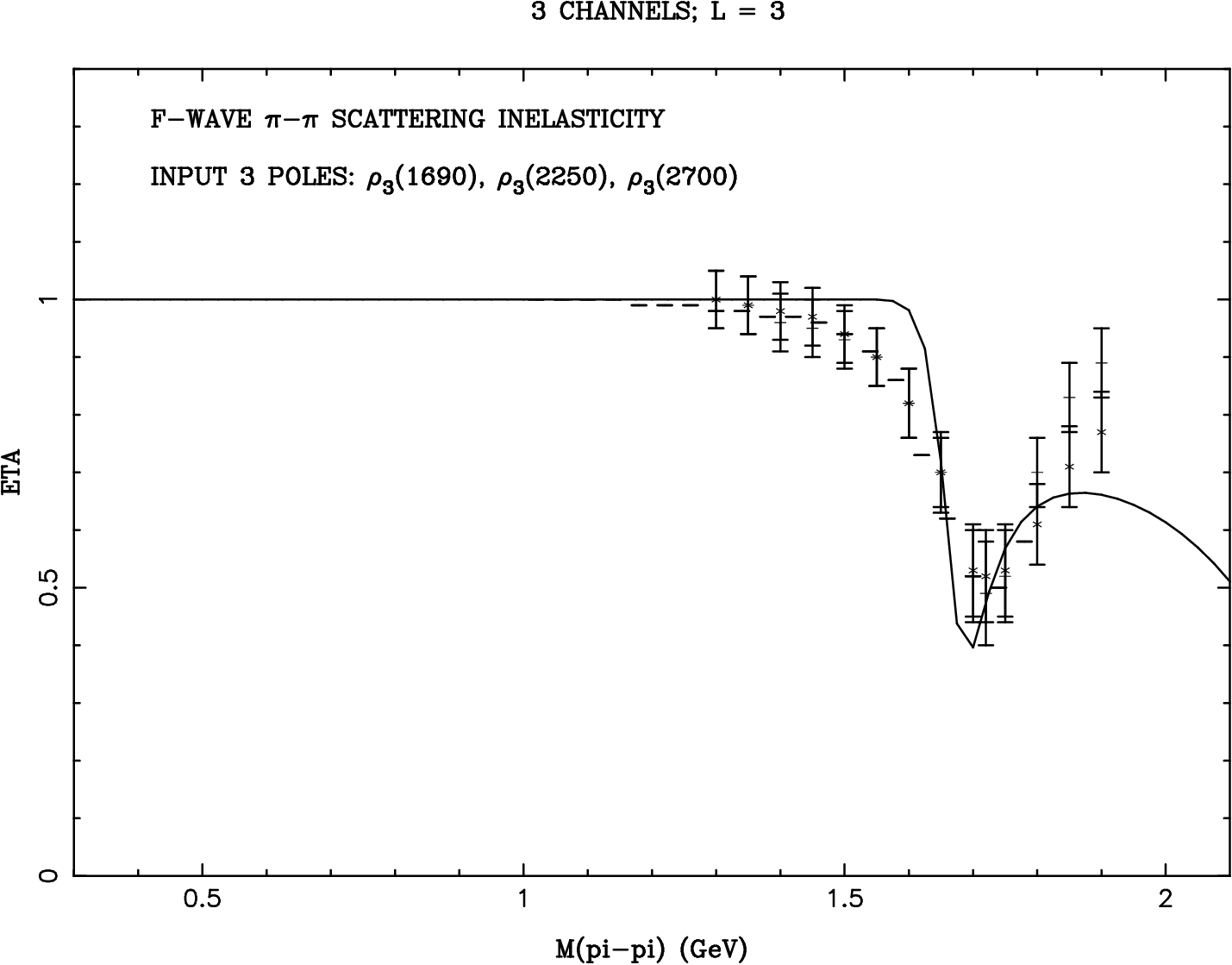,angle=-90,width=14cm
,bbllx=35pt,bblly=15pt,bburx=630pt,bbury=720pt,clip=}}  
\caption[]{
F-wave $\pi \pi$ inelasticities from the three-channel model, 
if poles are present for 
$\rho_3(1690)$, $\rho_3(2250)$ and $\rho_3(2700)$. 
Solid curve is the prediction for the $\it{l}$ = 3 parameter set from 
Table~\ref{table:LVAL3}. 
Data are from Refs.~\cite{Hya,Bug}.}
\label{fig:figeta3}
\end{figure}

For $J^{\pi}$ = $3^-$ there is one well established resonance 
$\rho_3$(1690), at (1.691, -0.080) GeV, 
and a brief mentioning of $\rho_3(2250)$. Since we are interested in
the phase shift below 2.0 GeV, the precise location of higher poles 
seems not very important. Within our model, having the essential 
ingredient of three channels and therefore nine parameters, a single 
resonance would allow too much freedom in the parameters. Just for 
convenience we impose therefore the condition that there are three
resonances, even for $J^{\pi}$ = $3^-$, at energies (1.691, -0.080), (2.250, -0.125), 
(2.700, -0.300) GeV. 
In that case it is straightforward to obtain a good fit to the 
experimental phase shift and inelasticity by the parameter set given 
in Table~\ref{table:LVAL3}. 
The model prediction for that set is given as the solid 
curves in Fig.~\ref{fig:figdel3} and Fig.~\ref{fig:figeta3}.

\section{DISCUSSION AND CONCLUSION}

In conclusion, we have constructed a three-channel model
which gives a reasonable description of the
S, P, D and F-wave  $\pi \pi$ scattering in the 0.3 -- 1.7 GeV region. 
Apart from the $\pi \pi$ and $\bar{K}K$ channels, the multi-pion 
channels are described as
effective $\rho \rho$ and $\omega \omega$ channels.  
As a result the model exhibits branch cuts at the $\bar{K}K$ and 
$\rho \rho (\omega \omega)$ thresholds. 
Subsequently we require the S-matrix to have poles at the three 
lowest known resonances for each angular momentum $J^{\pi}(I)$ 
as listed by the Particle Data Group~\cite{PDG}. 
The position of the resonances on the various complex momentum sheets 
reflects the decay properties of each resonance. 

As a by-product of this investigation 
one concludes that the $f_0(1500)$ resonance plays a 
role in S-wave $\pi^+ \pi^-$ scattering that is quite different from 
that of $f_0(980)$ and $f_0(1370)$. Also the S-wave and D-wave 
$\pi^+ \pi^-$ scattering data seem to be in better agreement with a 
value of $J$ = 0 for $f_J(1710)$ and at the same time 
in disagreement with a value of $J$ = 2. 
A further surprise for the $J^{\pi}(I)$ = $1^-(1)$ meson resonances 
is that not all three resonances $\rho (770)$, $\rho (1450)$, 
$\rho (1700)$, play a similar role in all three channels of this model.  

In order to establish how model dependent these conclusions are, it 
would be interesting to study the role of the various resonances off 
all angular momenta in other approaches~\cite{Kam1,Mul}. The present 
model can also be extremely useful to describe the final state 
interaction in reactions like 
$\bar{p}p \rightarrow \pi^+ \pi^-$~\cite{Bat}.

\acknowledgements

One of the authors (W.M.K.) is grateful to the Division de
Physique Th\'eorique, Institut de Physique Nucl\'eaire at Orsay for
its hospitality during his stay, when this work was initiated.
Both authors thank W.R. Gibbs, L. Le\'sniak, J. de Swart and R. Vinh Mau 
for helpful discussions.
This work is supported in part by NSF Grant No. PHYS-9504866.

\vspace*{0.5in}
\appendix{\bf APPENDIX A:}
{\bf Integrals for $X^l (s)$}

In order to evaluate the integrals for $X^l (p,s)$ of Eq.~(\ref{Xdef}), 
using Eq.~(\ref{gdef}) and~(\ref{Gdef}), i.e.

\begin{equation}
X^l (p,s) = \frac{2}{\pi} \int dp' p'^2 \frac{p'^{2l}} 
{(p'^2 + \beta^2)^{2 l + 2}} \frac{1}{p'^2 - p^2 - i \epsilon},
\end{equation}
for real p, we define a function $Z^l (p,s)$ by  

\begin{equation}
Z^l (p,s) = \frac{2}{\pi} \int dp' p'^2 \frac{1}{(p'^2 + \beta^2)^l} 
\frac{1}{p'^2 - p^2- i \epsilon},\label{Zdef} 
\end{equation}
also for real values of p. 
The values of $Z^l (p,s)$ for $l$ $>$ 1 can be determined, using the 
iteration method of the Appendix of Ref.~\cite{gib}, 

\begin{equation}
Z^{l+1} (p,s) = - \frac{1}{l} \frac{\partial}{\partial \beta^2} Z^l (p,s).
\label{Ziter} 
\end{equation}
All expressions for $Z^{l+1} (p,s)$ can then be determined from 

\begin{equation}
Z^1 (p,s) = \frac{1}{\beta - i p}.\label{Z1def} 
\end{equation}
The expression of Eq.~(\ref{Z1def}) for real p can again be 
analytically continued for complex p. 

The expressions for $X^l (p,s)$ follow from the binomial forms 

\begin{equation}
X^0 (p,s) = Z^2 (p,s),
\end{equation}

\begin{equation}
X^1 (p,s) = Z^3 (p,s) - \beta^2 Z^4 (p,s),
\end{equation}

\begin{equation}
X^2 (p,s) = Z^4 (p,s) - 2 \beta^2 Z^5 (p,s) + \beta^4 Z^6 (p,s),
\end{equation}

\begin{equation}
X^3 (p,s) = Z^5 (p,s) - 3 \beta^2 Z^5 (p,s) + 3 \beta^4 Z^7 (p,s) - \beta^6 Z^8 (p,s).
\end{equation}

\vspace*{0.5in}
\appendix{\bf APPENDIX B:}
{\bf Formulae with a second term in the $\pi\pi$ potential for the $l=0$ 
scattering.}

We give here explicit formulae for the Jost function and for the 
scattering length when we add, as in Ref. [2], a second term in the 
$\pi\pi$ channel for the $l=0$ scattering. The potential as well as 
the T-matrix of Eq.~(\ref{lipp}) are then 4x4 matrices. 
We choose the index 0 
to label this new term in the $\pi\pi$ potential of range $\beta_{0}$ and 
strength $\lambda_{00}$. We do not introduce any coupling between this 
term and the $\bar KK$ and $\rho\rho$ channels, i.e. $\lambda_{0i}=0$ 
for $i=2,3$ and furthermore $\lambda_{01}\equiv 0$.

The new Jost function can then be obtained in a similar way to the 
derivation performed in section II. One gets
\begin{equation}
	D(p_{1}p_{2},p_{3})=\left(1+\frac{\Lambda_{0}}
        {R_{0}}\right)Q(p_{2},p_{3})+
	\left[1+\frac{\Lambda_{0}}{R_{0}}
	\left(\frac{\beta_{1}-\beta_{0}}{\beta_{1}+\beta_{0}}\right)^{2}\right]
	\frac{P(p_{2},p_{3})}{R_{1}}.
	\label{B1}
\end{equation}
Here $\Lambda_{0}$ = $\lambda_{00}/(2\beta_{0}^3)$ and 
$R_{0}$ = $1/(2\beta_{0}^3X_{0})$ 
with, as in Eq.~(\ref{X0def}), 
\begin{equation}
	X_{0}=\frac{1}{2\beta_{0}(\beta_{0}-ip_{1})^2}.
	\label{B2}
\end{equation}
Eq.~(\ref{B1}) contains the expressions 
\begin{equation}
	Q(p_{2},p_{3})=1+\frac{\Lambda_{2}}{R_{2}}+\frac{\Lambda_{3}}{R_{3}}+
	\frac{\Lambda_{2}\Lambda_{3}-\Lambda_{23}^2}{R_{2}R_{3}},
	\label{B3}
\end{equation}
\begin{equation}
	P(p_{2},p_{3})=1+\frac{\Lambda_{1}\Lambda_{2}-\Lambda_{12}^2}{R_{2}}+
	\frac{\Lambda_{1}\Lambda_{3}-\Lambda_{13}^2}{R_{3}}+
	\frac{L_{123}}{R_{2}R_{3}},
	\label{B4}
\end{equation}
with
\begin{equation}
	L_{123}=2\Lambda_{12}\Lambda_{23}\Lambda_{13}-
	\Lambda_{13}^2\Lambda_{2}-\Lambda_{12}^2\Lambda_{3}-
        \Lambda_{23}^2\Lambda_{1}+\Lambda_{1}\Lambda_{2}\Lambda_{3}.
	\label{B5}
\end{equation}
It can be seen that if $\Lambda_{0}=0$, Eq.~(\ref{B1}) reduces to 
Eq.~(\ref{Jost}) as it should.

The $\pi\pi$ scattering length can be calculated as
\begin{equation}
	a_{\pi\pi}=\lim_{p_{1}\to 0}\frac{S_{11}-1}{2ip_{1}}=	
	\lim_{p_{1}\to 0}\frac{D(-p_{1},p_{2},p_{3})-D(p_{1},p_{2},p_{3})}
	{2 i p_1 D(p_{1},p_{2},p_{3})}.
	\label{B6}
\end{equation}
As $\lim_{p_{1}\to 0} A_{1}=1$, one obtains
\begin{equation}
	a_{\pi\pi}=\frac{-2P(p_{2},p_{3})/\beta_{1}-2\Lambda_{0}/\beta_{0}
	\left[Q(p_{2},p_{3})+P(p_{2},p_{3})\frac{(\beta_{1}-\beta_{0})^2}
        {\beta_{1}(\beta_{1}+\beta_{0})}\right]}
	{Q(p_{2},p_{3})(1+\Lambda_{0})+P(p_{2},p_{3})\left[1+\Lambda_{0}
        \left(\frac{\beta_{1}-\beta_{0}}
	{\beta_{1}+\beta_{0}}\right)^2\right]}.
	\label{B7}
\end{equation}
In Eq.~(\ref{B7}) $p_{2}^2=m_{1}^2-m_{2}^2$ and 
$p_{3}^2=m_{1}^2-m_{3}^2$.

\mediumtext

\begin{table}
\caption{
Relations of $\Lambda_{ij}$ to $\lambda_{ij}$ and $R_i$ to $X_i$ for 
angular momentum $\it{l}$.}
\begin{tabular}{ccccccccc}
$\it{l}$ & $\Lambda_{i}$ & $\Lambda_{ij}^2$ & $R_i$ \\
\tableline
0 & $\lambda_{ii}/(2 \beta_i^3)$ & $\lambda_{ij}^2/(4 \beta_i^3
\beta_j^3)$ & $ 1 / (2 \beta_i^3 X_i)$ \\
1 & $\lambda_{ii}/(16 \beta_i^5)$ & $\lambda_{ij}^2/(256 \beta_i^5
\beta_j^5) $ & $ 1 / (16 \beta_i^5 X_i)$ \\
2 & $\lambda_{ii}/(256 \beta_i^7)$ & $\lambda_{ij}^2/(256^2 \beta_i^7
\beta_j^7) $ & $ 1 / (256 \beta_i^7 X_i)$ \\
3  & $\lambda_{ii}/(2048 \beta_i^9)$ & $\lambda_{ij}^2/(2048^2
\beta_i^9 \beta_j^9) $ & $ 1 / (2048 \beta_i^9 X_i)$ \\
\end{tabular}
\label{table:PARA}
\end{table}

\begin{table}
\caption{Particle Data resonances considered in this work.}
\begin{tabular}{ccccccccc}
$\it{J}^{\pi} (\it{I})$ & $0^+ (0)$ & $1^- (1)$ &$2^+ (0)$ & $3^- (1)$ \\
\tableline
--- & $f_0(980)  $ &  $\rho(770) $ & $ f_2(1270) $ &  $\rho_3(1690)$\\
--- & $f_0(1370) $ &  $\rho(1450)$ & $ f_2(1525) $ &  $\rho_3(2250)$\\
--- & $f_0(1500) $ &  $\rho(1700)$ & $ f_2(1710)?$ &  $\rho_3(2700)$\\
--- & $f_0(1710)?$ &  $\rho(2150)$ & $ f_2(2010) $ &  ---\\
\end{tabular}
\label{table:resonances}
\end{table}

\begin{table}
\caption{Different model parameters for $\it{l}$ = 0, i.e. 
$\it{J}^{\pi} (\it{I})$ = $0^+(0)$ ($\beta$ in GeV), Figs. 1 -- 4.}
\begin{tabular}{ccccccccccc}
$\beta _0$& $\beta _1$& $\beta _2$& $\beta _3$& $\Lambda _0$& $\Lambda
_1$& $\Lambda_2$& $\Lambda_3$& $\Lambda_{12}$& $\Lambda_{13}$& $\Lambda_{23}$\\
\tableline
2.80& 0.85&  0.40&  1.75& -2.30& 4.05& -0.158& -1.300&  0.5523&  
0.0399& 0.5425\\
3.60& 1.35&  0.50&  0.25& -2.70& 3.95& -0.678& -5.561&  0.3059&  
0.3173& 2.0061\\
3.80& 1.25&  0.20&  1.00& -2.40& 3.80& -0.946& -1.575&  0.1689&
0.3487& 1.2793\\ 
2.80& 0.85&  0.30&  3.25& -2.30& 4.20& -0.124& -1.013&  0.5646&
0.3230& 0.4532\\
4.00& 1.25&  0.30&  0.50& -2.30& 3.70& -0.914& -2.656&  0.1333&
0.5098& 1.6397\\
\end{tabular}
\label{table:LVAL0}
\end{table}

\begin{table}
\caption{Different model parameters for $\it{l}$ = 2, i.e. 
$\it{J}^{\pi} (\it{I})$ = $2^+$(0) ($\beta$ in GeV), Figs. 5 -- 6.}
\begin{tabular}{ccccccccc}
$\beta _1$ & $\beta _2$& $\beta _3$ &$\Lambda _1$ & $\Lambda_2$&
$\Lambda_3$ & $\Lambda_{12}$ & $\Lambda_{13}$ & $\Lambda_{23}$\\
\tableline
2.400&   1.900&  0.750&  -0.210&  -0.252& -0.050&  0.0428&  0.1222& 0.0095\\
3.600&   0.300&  1.750&  -0.300&   1.596& -0.321&  0.1059&  0.0107& 0.0977\\
\end{tabular}
\label{table:LVAL2}
\end{table}

\begin{table}
\caption{Different model parameters for $\it{l}$ = 1, i.e. 
$\it{J}^{\pi} (\it{I})$ = $1^-$(1) ($\beta$ in GeV), Figs. 7 -- 8.}
\begin{tabular}{ccccccccc}
$\beta _1$ & $\beta _2$& $\beta _3$ &$\Lambda _1$ & $\Lambda_2$&
$\Lambda_3$ & $\Lambda_{12}$ & $\Lambda_{13}$ & $\Lambda_{23}$\\
\tableline
3.200&  15.500&  3.250&  -0.855&  -0.991& -1.051&  0.0035&  0.0843& 0.0074\\
2.600&   0.500&  0.750&  -0.560&   2.966& -1.457&  0.9756&  0.4386& 0.2775\\
\end{tabular}
\label{table:LVAL1}
\end{table}

\begin{table}
\caption{Model parameters for $\it{l}$ = 3, i.e. 
$\it{J}^{\pi} (\it{I})$ = $3^-(1)$ ($\beta$ in GeV), Figs. 9 -- 10.}
\begin{tabular}{ccccccccc}
$\beta _1$ & $\beta _2$& $\beta _3$ &$\Lambda _1$ & $\Lambda_2$&
$\Lambda_3$ & $\Lambda_{12}$ & $\Lambda_{13}$ & $\Lambda_{23}$\\
\tableline
3.800&   6.500&  0.750&  -0.170&  -0.188&  0.048&  0.0013&  0.0413& 0.0146\\
\end{tabular}
\label{table:LVAL3}
\end{table}

\end{document}